# A New Approach to Analyzing Patterns of Collaboration in Co-authorship Networks - Mesoscopic Analysis and Interpretation

Theresa Velden, Asif-ul Haque, and Carl Lagoze

*Computer and Information Science, Cornell University, Ithaca, U.S.A.*

tav6, ah344, cjl2 @ cornell.edu

**Abstract:** This paper focuses on methods to study patterns of collaboration in co-authorship networks at the mesoscopic level. We combine qualitative methods (participant interviews) with quantitative methods (network analysis) and demonstrate the application and value of our approach in a case study comparing three research fields in chemistry. A mesoscopic level of analysis means that in addition to the basic analytic unit of the individual researcher as node in a co-author network, we base our analysis on the observed modular structure of co-author networks. We interpret the clustering of authors into groups as bibliometric footprints of the basic collective units of knowledge production in a research specialty. We find two types of coauthor-linking patterns between author clusters that we interpret as representing two different forms of cooperative behavior, transfer-type connections due to career migrations or one-off services rendered, and stronger, dedicated inter-group collaboration. Hence the generic coauthor network of a research specialty can be understood as the overlay of two distinct types of cooperative networks between groups of authors publishing in a research specialty. We show how our analytic approach exposes field specific differences in the social organization of research.

## Introduction

Scientific fields differ in their intellectual and social organisation, and consequently in their communication practices (Whitley 2000, Cetina 1999, Fry 2007). Investigating these differences via ethnographic studies alone is insufficient because evidence is gathered from only a small, local fraction of scientists in a field. In contrast, a bibliometric approach that analyzes large sets of publication data provides access to aggregate behavioural patterns of a cross-section of scientists in a research field. However, because of the standardization of formal scientific publishing across science, the bibliometric approach employed alone may fail to uncover underlying, field specific differences in scientific communication practices. In our comparative research into field-specific scientific communication cultures we combine qualitative ethnographic field studies with structural analysis of publication networks. In this manner, we can use small-scale, nuanced evidence from field studies to inform our interpretation of large-scale publication networks and investigate to what extent field specific practices are reflected in structural features of publication networks. This evolves a tradition of close-up analysis of scientific networks and communication practices started by Crane's work (1972) on invisible colleges, and taken up more recently by Zuccala (2006).

Our specific concern is to understand the relevance and meaning of certain network features (such as co-author links and inter-citations) in the context of specific scientific fields, and to discover how they relate to concepts of community, and to communication practices that we observe in our field studies. The field specificity of global bibliometric indicators such as citations and journal impact factors (Moed 1985, Althouse 2008), the number of co-authors on a paper (Liberman1998), or the percentage of self-citations (Snyder 1998) is well known. However, the delineation of scientific fields, and hence the appropriate normalisation of such measures for comparisons across scientific fields remains problematic (Zitt 2005, Adams 2008). Recently, the characteristics of co-author networks as social networks have been highlighted (Kretschmer 1994, Newman 2001). They share with other social networks global topological properties such as small world-property, clustering, and assortative degree mixing as well as a long-tail degree distribution and a scaling law for the clustering coefficient (Sen 2006). In the recent surge of work on the analysis of complex networks, and in particular on clustering algorithms to extract the modular structure of real world networks, co-author networks rank among the most prominent examples (e.g. Radicchi 2004, Newman 2004, Palla 2005). Most work in this area does not focus on analysis of network features specific to a scientific field. Instead, existing work uses co-author networks as a way to demonstrate algorithmic advances by comparing global network properties of co-authorship networks to other types of 'real world' networks (such as transportation or biological networks).

Those works in information science and bibliometrics that investigate the characteristics of co-author networks in specific fields from a social network perspective tend to focus on global topological network properties, or on the position and ranking of individual actors within the network, see e.g. Kretschmer (2004), Acedo (2006), or Wagner (2005). See Liu (2005) for an exception that at least touches on the group level organization. Co-author network analyses at the global (or the individual author) level fail to recognize the team-based organization of research in most scientific fields. As Guimera et al. (2007)

point out for networks with a modular substructure, global properties fail to capture important structural and functional distinctions. To discover those one has to investigate the mesoscopic structure of networks that takes into account module interconnectivity and differences in connectivity patterns between nodes based on their structural position in the network. Accordingly the study presented in this paper focuses on the *mesoscopic level of analysis* - that is we analyze connectivity patterns between modules of closely interconnected authors in co-authorship networks in order to explore the field-specificity of community structures and communication patterns.

Early on in our research we found it problematic to interpret co-authorlinks between co-author groups as indicators of collaboration between those groups. We realised that in the weakly interconnected field that we were studying, co-authorship links between groups did not indicate direct inter-group collaboration but were just residues of the fact that individuals migrated between groups on their career path, e.g. from PhD student to postdoc, an observation made also by Nepusz (2008). In this paper we develop an approach to distinguish different types of connectedness that co-author links between groups represent such as inter-group collaboration, career migration or exchange of services or samples. To achieve this we match observations of structural features of co-author networks at the mesoscopic level with accounts of our field study participants on the underlying scenarios of interaction and coauthorship. We then proceed to compare collaboration patterns of three research specialties in chemistry as they are reflected by the mesoscopic structure of their co-author networks.

A distinguishing aspect of our research is the manner in which qualitative and quantitative approaches are closely interlinked. The study presented here is exemplary for the way in which both approaches interact: qualitative understanding of research practices and context help us not only to evaluate quantitative outcomes, but also to further evolve our quantitative methods, and quantitative results guide our attention for further qualitative study. Take for example the issue of analyzing the community structure of social networks, which is an important aspect of our research into communication practices. As described further below, a wide variety of clustering algorithms exist that will all deliver quite different partitions of a network. To decide whether author groupings obtained from clustering of a co-author network provide us with a meaningful research unit for the analysis of scientific communication and interaction patterns of a research field, we need validation and interpretation of those clusters in the context of this specific field (Caruana 2006, Schaeffer 2007). Given such a clustering we can then look at the network of clusters that shows the interactions between author groups and direct our qualitative research efforts at further understanding those interaction patterns – that is, to uncover those motivations and processes that underlie the structural patterns (Lievrouw 1990, Zuccala 2006).

**Methodology**

To develop a deep qualitative understanding of communication practices in different fields of chemistry we are conducting ethnographic field studies in selected research groups in Europe and the U.S.A. We will call these groups *seed groups* in the remainder of this paper[i]. A precondition for selecting a group for our field study is that it is an internationally recognized player in some specialised field of research, below the sub-discipline level. The publication data sets used in this study represent those research specialties in which our seed groups are actively engaged. As a result, for each data set we have access to informants with whom we can check the interpretation and significance of features that we detect in our network analysis.

**Data**

We have obtained data from Web of Science (WoS, Thomson Reuters) using a lexical query (see appendix) to capture the literature published in small specialities within synthetic chemistry (Field B) and physical chemistry (Fields A and C)[ii] over a ~20 year time period. For the construction of the co-author networks we have omitted those authors that have published only a single paper within the data set. The original number of authors and the reduced number of authors in the co-author networks are given in table 1.

|  | **Field A** | **Field B** | **Field C** |
|---|---|---|---|
| Sub-discipline | Physical chemistry | Synthetic chemistry | Physical chemistry |
| Time span | 1991-2008 | 1991-2008 | 1987-2008 |
| Average # of authors per paper (median) | 3.8 (3) | 3.3 (3) | 3.8 (3) |
| # authors | 81,606 | 20,643 | 47,323 |
| # authors (reduced) | 33,102 | 7,465 | 18,419 |
| # authors in giant (relative size) | 31,480 (95.1%) | 6,668 (89.3%) | 17,250 (93.7%) |
| # clusters in giant | 2005 | 578 | 1191 |
| Average # of authors per cluster (median) | 16.4 (11) | 12.1 (9) | 15.2 (10) |

**Table 1. Co-author Networks**

We conduct a crude temporal analysis of network growth by dividing the time period into time slices of roughly equal length (see figure 1). It shows an increasing rate of new authors entering the fields. It further shows that in all three fields the relative size of the giant component converges to values about 90% and higher[iii] (see table 1), indicating that the 18-22 years time window is sufficient for most basic relations between actors to have developed (Newman 2001a, Barabasi 2002).

The field studies helped us to understand the research activities our seed labs were involved in, the social organization of the research groups, and experiences people had made in collaborations with other groups. Field visits with observations and interviewing of group members lasted between 4-6 weeks per group. Altogether we visited 5 different groups. Further, we acquired specific feedback to help interpret the structural features we were extracting from the coauthor networks. We conducted such interviews with the PI or another senior researchers in a group. The development of the analysis presented here is based on feedback interviews with 5 different researchers, amounting to about 15 hours of conversation, partially audio recorded, sometimes documented by extensive note taking.

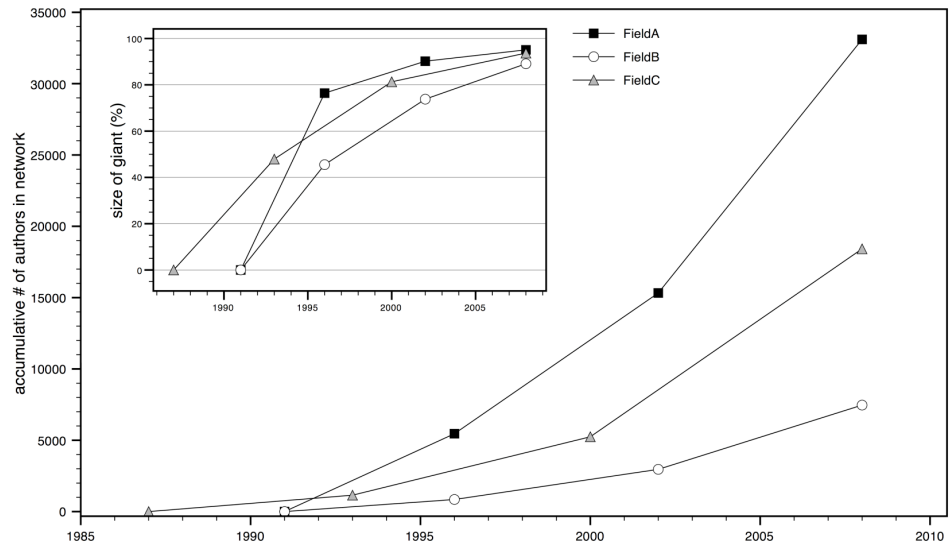

**Figure 1. Accumulative temporal evolution of co-author numbers in network (large diagram) and of relative size of giant component (inset).**

**Procedure**

The nature of the node 'communities' that are extracted from a network depends on the clustering algorithm that is used (Fortunato 2007). Some clustering procedures are hierarchical by design, so that they offer a hierarchy of groupings, starting from single nodes to the total set. Newman (2004) shows how his non-hierarchical algorithm can still be used to repeatedly drill down into a large, 56,000 author data set, first detecting clusters at the level of entire fields (such as high-energy physics, or condensed matter physics), then smaller specialities, and eventually after a fourth iteration, clusters corresponding to research groups (such as his own with 28 members). Other algorithms have a very restrictive definition of clusters as very tight groupings of nodes, e.g. the concept of k-cliques in Palla et al. (2005), and therefore produce clusters with little variation of group internal structures.

We use an information-theoretic clustering algorithm[iv] for undirected, integer-valued networks (Rosvall 2007) to partition the co-author network into clusters of closely interconnected co-authors. We have chosen this algorithm because we have found that at least for the few seed groups about which we have some in-depth knowledge from our field studies, it extracts very well what Seglen (2000) has called *'functional research groups*' – that is, basic research collectives that not only contain a collocated group of researchers in a laboratory led by a principal investigator (PI), but also closely cooperating domestic or international colleagues and visiting scientists.

We then proceed as follows. First we investigate the basic components of the mesoscopic structure of the clustered co-author networks and their interpretation. The basic components are the clusters, and the linkages between the clusters. Then we compare properties of these components for the three fields in our case study.

1. We extract the giant component of the co-author networks. Then we use pajek (Batagelj 2003) to extract and visualize the cluster that includes the PIs of our seed groups, as well as some of the clusters in their neighbourhood. We validate the interpretation of clusters in interviews with our informants.
2. Using standard network centralization indices (degree, closeness, betweenness; Freeman 1978) calculated with pajek, we compare the internal structure of selected clusters. We further explore the composition of clusters using the approach by Guimera et al. (2007) to distinguish different types of nodes based on their cluster-internal and cluster-external links. The classification distinguishes seven node roles. It distinguishes between hub-nodes and non-hub nodes based on the number of cluster internal links a node has. Further subtypes of hubs and non-hubs are defined based on the distribution of their external links to other clusters (see method section in appendix for details).
3. For analysis of the linking patterns between clusters we focus on the neighbourhood of a seed group within the original giant component of the *author-level network* – that is, we extract the sub-network that contains all authors of the seed-group and all authors of those clusters that are linked by at least one co-author link to the cluster of the seed group. We visualize these sub-networks in pajek and inspect the connection patterns between the seed cluster and its neighbours.
4. We then ask our informants from the seed groups to review their neighbourhood network and to tell us about their scientific relationship to the neighbouring clusters, and how the co-author links have come about. In those interviews typical scenarios emerged that we could match with typical linking patterns as described below in the results section. They are the foundation for the classification of between-cluster connections in the next step.
5. Based on our observations in the previous step, we classify links between clusters into two types and build from each type a *cluster-level network* where nodes represent clusters: we regard two clusters as connected by a *transfer* type link if in the underlying co-author network the two cluster are no longer connected by co-author links when (hypothetically) one or two author nodes from the network are removed. In all other cases we classify between-cluster linking as *collaboration*. In both cases we assign as weight to the cluster-level link the sum of the weights of the underlying co-author links.
6. We conclude our investigation by comparing systematically the empirical data for three research fields with regard to the properties of their population of clusters and the properties of their transfer and collaboration networks.

# Results

## 1. Basic Components of the Mesoscopic Structure of Co-author Networks

*1.1 Interpretation of Clusters*

In interviews with the PIs of the seed groups we validated that their respective cluster extracted from the co-author network captures their immediate, collocated research groups (researchers of various seniority levels from PhD students over postdocs to deputies or subgroup leaders) plus relevant external cooperation partners (from their own institution, or national and international cooperations). Occasionally the PIs were slightly surprised that a certain individual was subsumed into their cluster and not independently represented. This particularly seems to be the case for those individuals that have a strong institutional identity distinct from the academic group of the PI – as is the case for research group leaders from industrial companies. Note that our data sets capture only publication activity in very specialised areas of research – many PIs and their groups in our field study engage in several distinct research specialties. Consequently the representation of research groups by co-author clusters in this study is only partial, as it includes only those group members that are involved in the specific research specialty that we targeted when defining our data set.

The seed groups of the PI-led research groups show a centralized hierarchical structure with the most productive author in the centre, see for example the cluster that corresponds to a seed lab in field B, depicted at the top in figure 2. The network centrality indexes of degree (0.88), closeness (0.91), and betweenness (0.86) for this cluster indicate a high level of centralization, not far from 1, the highest possible value obtained e.g. for a perfectly star-shaped network. We further find that some of the clusters show a quite different internal structure such as the examples from field C depicted in the bottom row of figure 2. They include several central nodes none of which dominates the entire cluster. The clusters are still hierarchically organized in the sense that some nodes are clearly larger than most other nodes, and that those smaller nodes are mainly linked to the large nodes. Their centralization indices range between 0.21 and 0.50; they are substantially lower than those for the seed group cluster.

The node role analysis underlines this observation of differences in cluster organization. Based on the internal linking structure of a cluster it distinguishes hub-nodes and non-hub nodes. The PI-led group depicted in figure 2 is characterized by having only one single hub-type node, whereas the other two clusters show several hub-type nodes. From investigation of institutional affiliations given in the underlying WoS records and participant feedback, we identify one of these latter clusters as an international collaboration network, and the other as a network of closely cooperating colleagues from a major research institute in this research specialty.

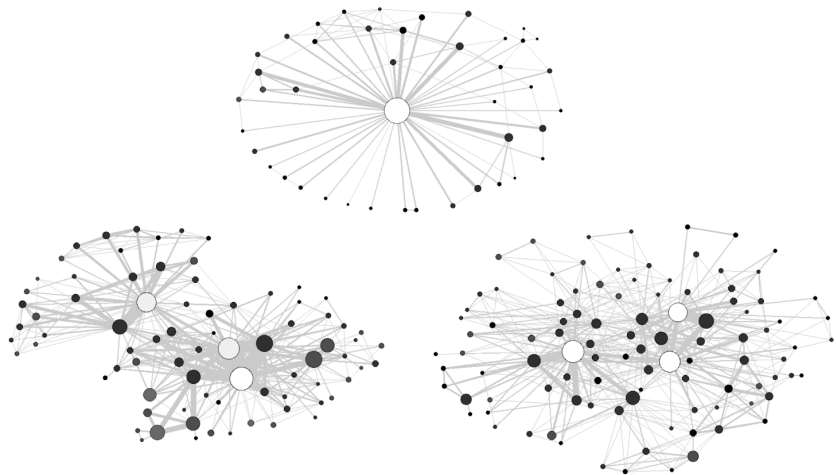

**Figure 2. top: PI-led laboratory group; bottom left: international research network (predominantly Japanese, Korean, and French); bottom right: cooperative network of a research Institute. Nodes represent authors; node sizes the number of papers they have authored, fill color: 'hub nodes' are highlighted by white fill colour, 'non-hub' type nodes are filled by shades of dark grey. Links represent co-author links, the thickness of links the number of co-author events.**

*1.2 Between-Cluster Linking Patterns*

On visual inspection of the linking patterns between the seed clusters and their neighbour clusters we identify two major types of structural patterns: connections where a single author connects two clusters, and connections where substantial fractions of the nodes of both clusters are involved. Figure 3 shows a variety of connection patterns between the seed group cluster in field C and a subset of its neighbour clusters. We distinguish two structural subtypes of the connection type where a single author connects two clusters: *1-1* connections and *1-m(any)*. We also observe a few cases where two repeated instances of *1-m* connections seem to constitute the only linkage between two clusters. Hence a pair of authors connects these clusters: we denote those connections as *2x (1-1)* or *2x (1-m)*. We also distinguish two structural subtypes of the *m-m* connections that involve many authors on both sides: (A) those connection patterns that include direct PI-PI co-author links and seem to have a rather deep penetration into the cluster, and (B) those that involve only interaction with a subgroup of the seed cluster but not (or only minimally) the PI.

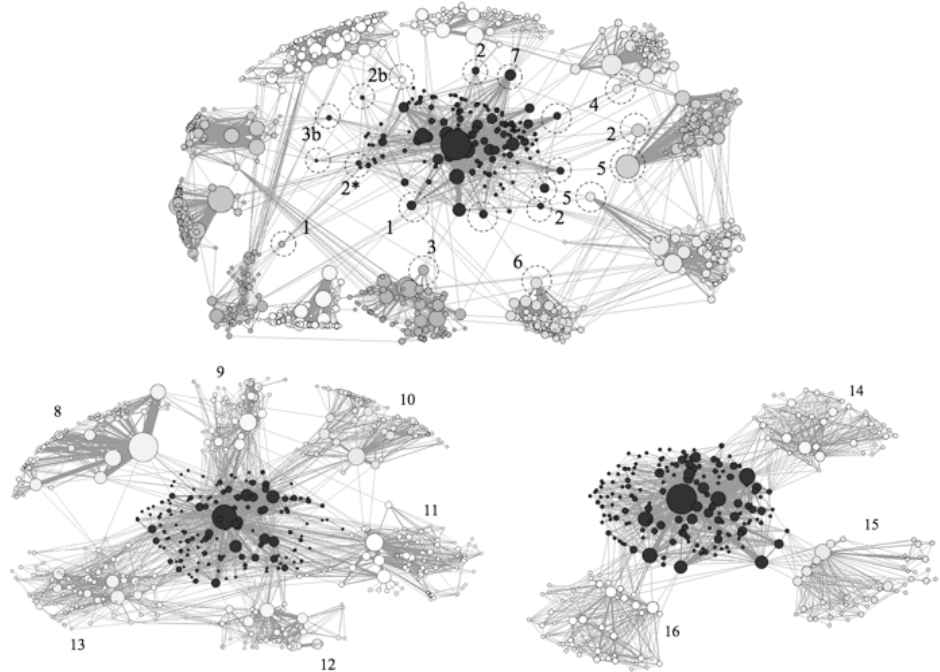

**Figure 3. Selected instances of how the seed group cluster in field C connects to neighbouring clusters. A node represents an author, node size number of publications, node shade cluster membership, a link co-authorship, and link thickness number of joint publications. Top: 1-1 and 1-m connections (dotted circles highlight involved nodes). Bottom: m-m type connections, left: with direct PI-to-PI links, right: without or with weak PI-PI link. The numbers given refer to the description of the underlying collaboration scenario derived from interviews (see list in text).**

In the list below, we match these connection patterns to the commentary of one of our informants on the real world scenarios underlying those co-author links. Numbers in parentheses identify the corresponding pattern in the networks in figure 3. These examples are taken from the seed group in field C – similar patterns have been observed and validated for the other fields.

*1-m connections* (one single node connects two clusters)
- visiting scientist with links home into national research network (1)

- career migration (2)
- repeated instances of career migration (2b)
- career migration of now closely collaborating colleague (2*)
- 1-off commissioned work (3)
- 1-off cooperations on independent topics (3b)
- funded project collaboration of subgroup leader (7)

*1-1 connections* (two nodes, one from each cluster, are exclusively linked)
- 'unauthorized' collaboration by postdocs (4)
- provision of synthesis samples (5)
- exclusive cooperation by closely collaborating colleague (6)

*m-m connections A* (interlinked node groups from each cluster participate in the connection including the PIs)

- intensive thematic and methodological international (Swiss) cooperation for since PI's postdoc time (8)
- very broad, thematic based, many faceted international (UK) cooperation (includes career migration and temporal exchange of postdocs and PhD students) (9)
- many-faceted methodological international (Dutch) cooperation (10)
- cooperation on many topics due to complementary knowledge, joint PhD students and exchange of staff, many funded projects (11)
- PI colleague at same institute - intensive topical and methodological cooperation; much stronger informal cooperation than formal in form of publications (12)
- multi-faceted cooperation with national institute, exchange of senior staff who bring along their networks (13)

*m-m connections B* (interlinked node groups from each cluster participate in the connection with no or minimal PI-PI links)

- Thematic cooperation, nurtured by EC funding, no exchange of PhD students; people come for measurements and leave again; disciplinary very distant (electrical engineers) (14)
- Seed group conducts measurement services to this group, only subgroup of instrument involved, institutionally supported by partner agreement between institutions, PI former postdoc in seed group (15)
- Russian institute, seed group is partially fused with this institute as several members are partially funded by seed group, repeated exchange of coworkers (16)

## 2. Case Study: Comparison of Mesoscopic Structure for Three Research Specialties

The purpose of this case study of three specialty fields within chemistry is to demonstrate the kind of comparative analysis our approach enables and the sensitivity of this approach for uncovering differences in the collective organization of scientific research specialties.

*2.1 Properties of Cluster Populations*

As reported in table 1, the clustering algorithm extracts from the giant component of field A 2005 clusters, from field B 578 clusters, and from field C 1191 clusters. The following properties refer to these cluster populations that make up the giant component of the network in each field.

For all fields, the size of the clusters in terms of number of authors correlates with the total number of publications co-authored by any of a cluster's authors as can be gleaned from the scatter plot for field A in figure 4. In first approximation this can be described as a linear correlation with Pearson correlation coefficients of 0.87 (field A), 0.94 (field B), and 0.90 (field C), respectively, indicating that for field B the correlation between cluster size and published papers is strongest. This finding is in agreement with previous findings, e.g. a study on research productivity of microbiology research groups in Norway (Seglen 2000).

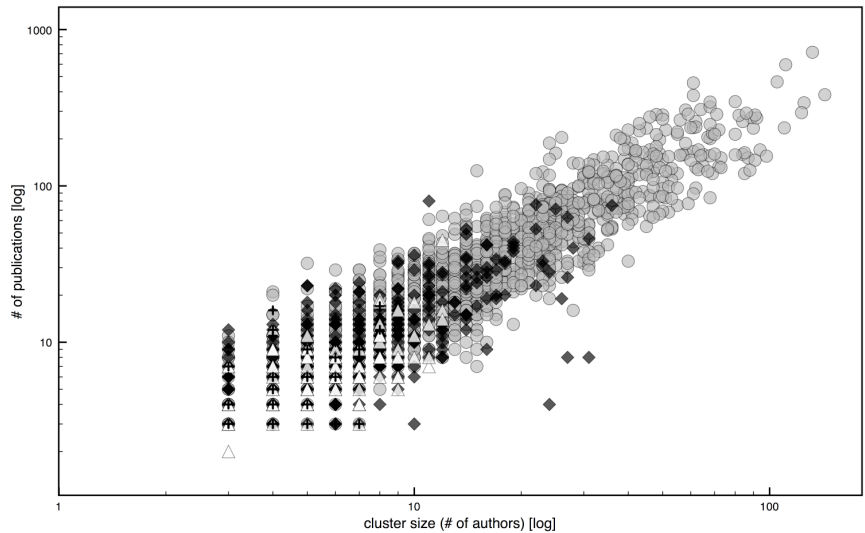

**Figure 4. Age of clusters highlighted in log-log scatter plot of size and publication output of clusters for field A. Each symbol represents a cluster. Shape and colour indicate membership in cluster age cohort: grey circle =** *continuous***, black diamond =** *recent***, white triangle =** *new***, black cross =** *extinct***. The definition of age cohorts is given in the text.**

For further analysis of the associations between various cluster properties and how they vary across fields, we group the clusters by size as either *small* ($n \leq 10$ authors), *medium* ($10 < n \leq 40$), or *large* ($40 < n$). To investigate the temporal composition of the cluster population for each field we divide the time period covered by our data sets into three slices of approximately equal length. We distinguish four age cohorts of clusters based on the publishing activity of their authors during these time slices: *continuous* – cluster authors have published during all three time slices; *recent*: publishing activity only during the latter two time slices; *new*: publishing activity only during the last time slice; *extinct*: active in first and/or second time slice, but not in most recent time slice.

Average cluster sizes are smaller in field B, than in fields A and C (see table 1). The medians point to a highly right skewed distribution in particular for fields A and C (a diagram showing the numeric distribution of clusters sizes for all three fields is provided in section 1 of the supplementary material). The largest cluster in field B has 91 authors, whereas each of the fields A and C has seven very large clusters with between 100 and 159 authors. In terms of the categorical cluster size variable defined above we find that field B has the largest proportion of small clusters (~59%), and it has fewer medium and large clusters than fields A and C (see row 1 of figure 5).

Figure 4 suggests for field A an association of cluster size with membership in a specific age cohort; scatter plots for fields B and C (not shown here) display the same trend. Indeed, taking the data of fields A, B, and C together, we find a statistically significant association between the categorical variables of cluster size and cluster age. Whereas overall 50.3% of clusters are small, for the age cohort of continuous clusters the proportion of small clusters decreases to 35.5%.

As depicted in row 2 of figure 5, all fields have only a small percentage of extinct clusters (3-4%), and the most numerous age cohort are the continuous clusters (48-74%), the second largest the recent clusters (19-40%), and the smallest surviving cohort are the new comers (5-11%). In comparison, field B has relatively more recent clusters than fields A and C, field C has relatively more new clusters than field A, and B, and field A has the largest proportion of continuing clusters. Note that these statements refer to a clustering obtained post hoc from data accumulated over the entire time period. Hence these observations do not translate directly to growth in terms of the formation of new clusters. If a scientist e.g. after a postdoctoral position with a group moves on, and over time during her or his career becomes the seed of a new group that is identified by our post-hoc clustering, this new cluster will still be categorized as continuous and not new, because at least one of its members (the former postdoc) has been publishing in the field all along. Nor do these statements on the relative sizes of cluster age cohorts translate directly to growth in terms of the entrance of new authors into the field. This can be seen when comparing with figure 1 that shows that all three fields have gained the largest portion of new authors during the most recent time slice whereas the largest portion of clusters are continuing clusters.

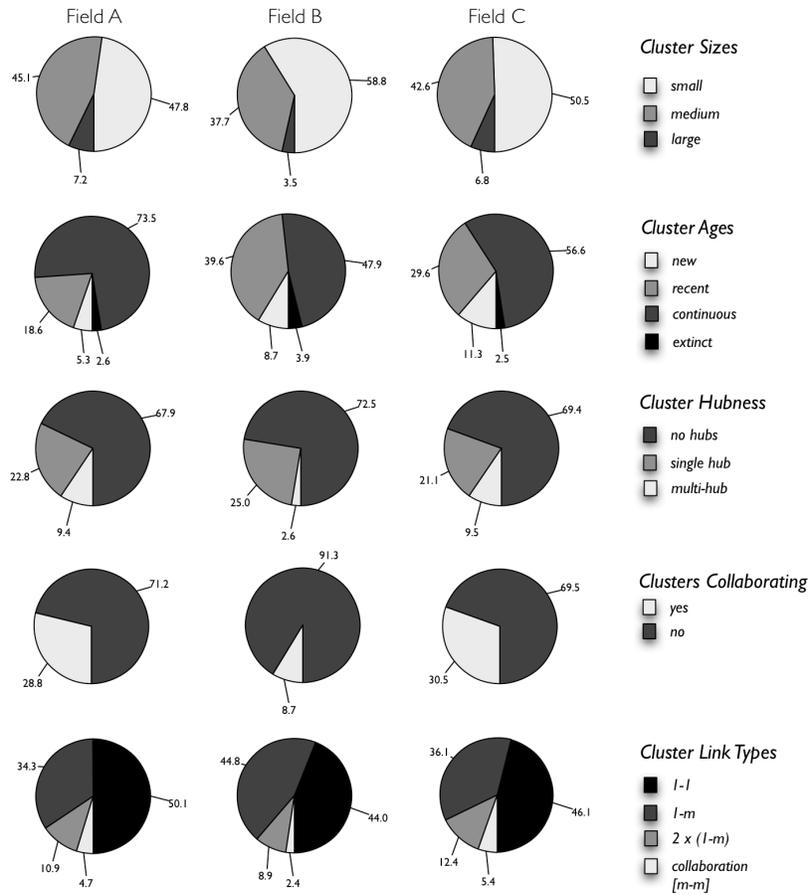

**Figure 5. For fields A, B, and C: properties of clusters in the giant component of the co-author network[1].**

Applying Guimera et al's node role classification scheme to the giant components of the clustered networks, we can distinguish between clusters with a single hub node, clusters with multiple hub nodes, and those without any hub nodes. The scatter plot for field A in figure 6 indicates that the hubness of clusters is associated with cluster size – the larger the clusters, the more hubs they contain (a trend that holds for all three fields). If we compare the hubness of clusters between fields, we see that there are slightly more hubless clusters in Field B, than in fields A, and C, and more multi-hub clusters in fields A and C (see row 3 in figure 5). Field B has the relatively highest proportion of single-hub clusters of the three fields. Both these observations, dependence of cluster hubness on cluster size and field-dependence of cluster hubness, are statistically significant results (see sections 2 and 3 of the supplementary material for details).

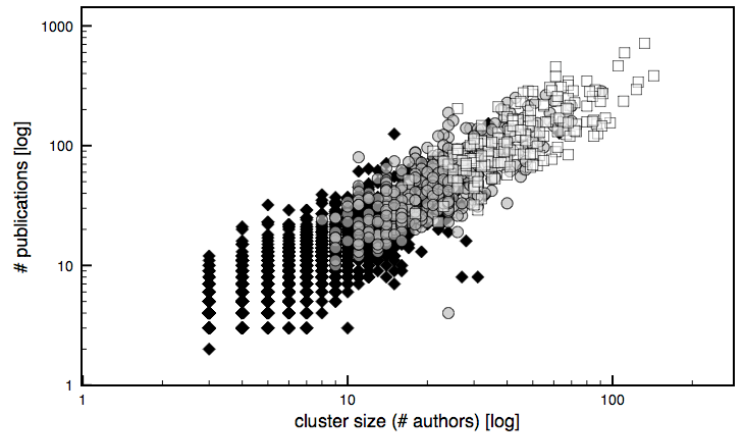

**Figure 6. Hubness of clusters highlighted in log-log scatter plot of size and publication output for clusters for field A. Each symbol represents a cluster. Shape and colour indicate hubness of a cluster: white square: multi-hub, grey circle = single hub, black diamond = no hub.**

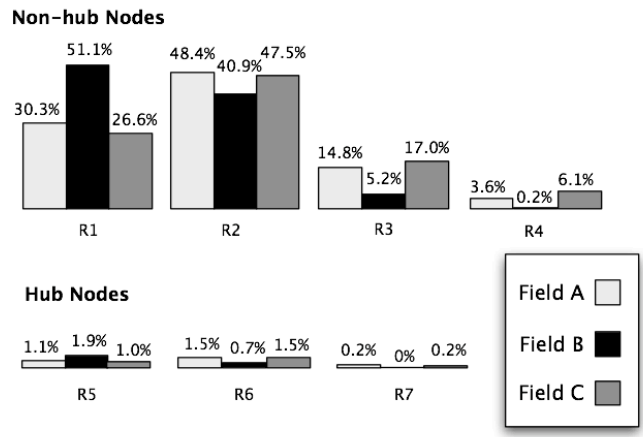

**Figure 7. Node role type distribution for fields A, B, and C. R1=ultra-peripheral node, R2=peripheral node, R3=satellite connector node, R4=kinless node, R5=provincial hub, R6=connector hub, R7=global hub.**

The distribution of the seven different node role types as defined by Guimera et al. is given in figure 7 for all three fields. The node roles of field B seem to be more 'shifted' towards node types with less external links than for fields A and C; so B has more ultra-peripheral (R1) than peripheral (R2) type nodes, whereas A, and C have more peripheral type nodes than ultra-peripheral type nodes. B has about 5% satellite connector (R3) and kinless nodes (R4) whereas fields A and C have around 20% of those node role types. B has more provincial hubs (R5) than connector hubs (R6), whereas A and C have more connector hubs than provincial hubs. Finally B has only one global hub (R7), whereas A and C have quite a few (in absolute numbers, 91 and 34 respectively).

*2.2 Collaboration and Transfer Networks*

Following our classification of between-cluster links into *transfer* type links (meaning that the hypothetical removal of one or two nodes results in the two clusters no longer being connected) and *collaboration* type links (all other cases), we obtain for each field two cluster-level networks that differ greatly in size: a transfer network, and a much smaller collaboration network. Of all between-cluster links, only a very small percentage qualifies as collaboration type links: for fields A and C these are 4.7% and 5.4% respectively, about twice as many as in field B with 2.4%, see row 5 in figure 5. Whereas for all three fields almost all clusters of the giant component are part of the transfer network, the fraction of clusters that are involved in the collaboration network is considerably smaller: for field A 28.8% (= 560 clusters), for field C 30.5% (= 348 clusters), and for field B 8.7% (= 49 clusters); see row 4 in figure 5. The collaboration networks of fields A and C have a giant component, whereas the collaboration network of field B is fragmented into several unconnected components of similar size.

To get an idea of the local densities of the collaboration networks, we look at the degree distribution for the cluster nodes in the collaboration networks Whereas the median degree of the clusters is 2 for all three fields, the average node degree for fields A and C is higher (3.3 and 3.1 respectively) than for field B (2.2), indicating a more strongly right-skewed distribution. Indeed, the highest cluster degree in field B is 7, whereas in fields A and C, 10.6% (37) clusters, respectively 6.3% (126) clusters have degrees higher than 7, going up to a degree of more than a hundred for one of the clusters in field C. On inspection of the names of the most productive authors in the high-degree clusters with more than 20 collaboration links we find almost exclusively common Chinese and Korean names, an observation we will come back to in the discussion section.

Fields A and C, show a higher local density of their transfer networks as well: the medians of cluster degrees are 15, and 11, whereas the median for field B is 6. The average cluster degrees for fields A and C are about 29 and 21, whereas for field B the average cluster degree in the transfer network is only about 8.

Subtracting the background of transfer type relations from actual inter-group collaboration between co-author clusters allows us to study with greater precision the properties of the collaboration network in a field. For example, an interesting question is to what extent geographical proximity is correlated with inter-group collaboration in a field. We present in figure 7 the collaboration networks of the three fields with colours highlighting their geographical affiliation (how we derive the geographical labeling is described at the end of Appendix A).

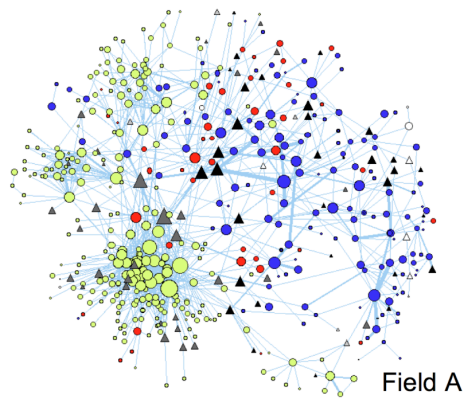

Field A

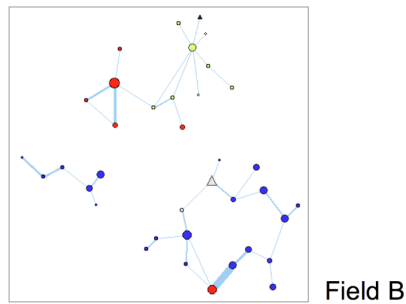

Field B

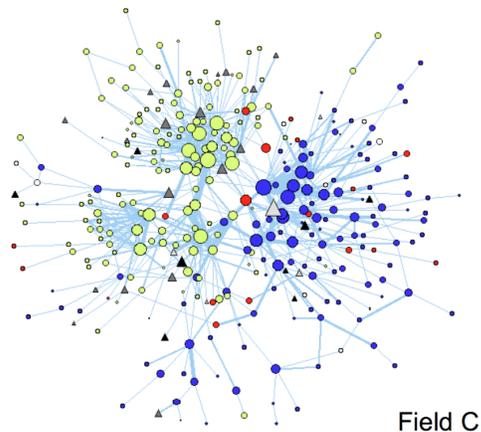

Field C

**Figure 8. We represent here the largest components of the cluster-level collaboration networks for the three fields. For simplicity we omit the few small dyads and triads that also exist. Nodes represent clusters of co-authors, lines between clusters represent co-author relationships of type *collaboration*, and node shapes and colours indicate geographical affiliation of co-author clusters: green circles - Asia, blue circles - Europe, red circles – North America; light grey triangles - Asia-Europe, dark grey triangles - Asia-North America, and black triangles - Europe-North America; all other continental affiliations are represented by white circles or triangles. For visualization we use pajek, and the network layouts are generated in pajek using the Kamada-Kawai, and Fruchterman Reingold algorithms. We have manually moved some central nodes to reduce overlap, and outer nodes to increase compactness of the representation.**

## Discussion

The analysis of mesoscopic network features provides us with insight into the underlying social configurations and processes of scientific collaboration and cooperative behaviour, and generates new research questions that we will highlight in the following sections.

### Interpretation of Basic Mesoscopic Features of Co-author Networks

We observe that the basic collective units performing research in the specialties that we are studying differ substantially in structure and composition. The uni-centred hierarchical footprint of a PI-led group (corresponding to the functional research groups Seglen (2000) identified in his case study on microbiology) contrasts with the multi-centred research clusters. These latter clusters have lower centralization indices and contain multiple hub-type nodes. They correspond to multiple-PI collaborations representing intra-institute, national, or international research networks. We conclude from this observation that a variety of social configurations and collaborative working modes contribute to research in a specialty.

Our structural analysis of between-cluster co-author links and how they match to underlying real world scenarios points to a fundamental difference between 1-1, 1-m or 2x(1-m) co-authorship links on the one hand, and m-m co-authorship links on the other hand. These basic types imply different forms of interaction and collaboration. The former are footprints of people moving between labs and of one-off cooperations involving sample exchange or measurement services rendered[v]. According to our informants' accounts, migration events can take different forms that are more or less strategically influenced by the interests of the home lab: 1) reciprocal exchange of students or young researchers between complementary groups, 2) targeted acquisition of training in a particular domain or skill that a young researcher brings back to his home lab when returning after a postdoc stint at another lab, or 3) career migration where the individual student or young researcher chooses rather independently what group to join next to evolve his or her career – personal recommendations of the current PI and his or her network of acquaintances, as well as the reputation and resources of the new lab influence such decisions. Hence this type of inter-group connections that we have termed *'transfer links'* constitutes a group-level network of exchange of people, skill, material, and measurement services within a specialty.

The m-m connection patterns on the other hand correspond to substantial inter-group collaborations. They are substantial in the sense that they are recognized by the leaders of our seed groups as inter-group collaborations worth mentioning when reflecting on their scientific career and evolution of research interests. Such inter-group collaborations include several group members, result in several publications, and extend over some extended time period. Distinguishing between those two basic structural types enables us to investigate and compare collaboration practices in scientific fields in these two different dimensions.

We expect that the rigorous structural distinction between the two basic types will occasionally lead to an erroneous classification of underlying collaboration scenarios. For example, an inter-group collaboration may expose the misleading structural form of a 2x(1-m) transfer link in a situation where an intensive inter-group collaboration includes only two members from one group; for a small group, two authors may correspond to a substantial fraction of that group (a realistic scenario would be a small independent theory group collaborating with a larger experimental group). We suspect that this kind of classification error is negligible: the distribution of linking types depicted in row 5 of figure 5 indicates that in the fields we are studying the 2x(1-m) structural pattern occurs only in about 10% of the cases overall. Furthermore, the large majority of cases that we encountered in the analysis of the neighbourhood networks of our seed groups represented traces of two separate migration events between the clusters, and was not the footprint of inter-group collaboration.

Based on these interpretations of basic mesoscopic features of co-author networks we will proceed with a comparative study of collaboration patterns reflected by co-author networks of three research specialties. We will first discuss generic findings that hold for all three fields, before we address the question what we learn about differences between the fields.

**Mesoscopic structure of co-author networks in chemistry: generic findings**

The pie chart table in figure 5 gives a systematic overview of the properties of the cluster populations in the three fields. There are a number of generic findings that hold for all three fields. First, cluster size is associated with cluster age, generally the older the cluster the larger the cluster. For the most part, this is an artefact of an accumulative data set, since we do not delete nodes or links from the network after a certain period of 'inactivity'. Hence especially for the older clusters, we do not capture the clusters as instantiated at any given point in time, but include a halo of authors who have in the meantime left the group and perhaps the field, or even science. So increasing the temporal resolution and studying the temporal evolution of clusters should give better insight into what constitutes genuine growth, and a genuinely large cluster in a research specialty at a given time, and whether there are any additional selection effects at play that make small clusters more likely to discontinue than large ones. A further line of inquiry for future work is to look at the substantial portion of continuous clusters (between 30-45% depending on the field) that remain small although they have been active over the entire time period.

Second, the hubness of clusters is also associated with cluster size; the larger the cluster, the more likely it is that it will include a hub node, or for very large clusters even several hub nodes. Whereas the interpretation of a single-hub cluster as a research group led by a PI seems straightforward, the question arises about what social configurations and collaborative working modes underlie the small non-hub clusters and the large multi-hub clusters. Our initial observation of multi-centre structures (see figure 2) that correspond to a specialised research institute and a closely collaborating international research network exemplify possible scenarios for multi-hub configurations of clusters. Hence, we take the fraction of multi-hub clusters in a field as a first, rough quantitative estimate of the frequency of either institutional or international multi-centre networks in the three fields: 2.6% in Field B, and 9.4%, and 9.5% in fields A and B, respectively. We speculate that small, hubless clusters on the other hand may represent small-scale, informal collaborations of equal ranked researchers. Alternatively, they may correspond to the early co-author footprint of groups newly entering the field. Such embryonic footprints would be too weak to reflect the underlying social hierarchy in seniority and productivity. However, the fact that most of these small hubless clusters are not new, but are either continuous or recent (see section 4 of the supplemental material for details), seems to devalue the latter argument somewhat. Therefore, further research is needed to understand the composition and role of those entities in a research specialty that are represented by small hubless clusters in the coauthor network.

Finally, we find that the clusters in the collaboration networks tend to be larger and older clusters (see section 5 of the supplementary material for details). In our interviews participants indicated a number of preconditions for successful inter-group collaborations, emphasizing that the 'inter-personal chemistry' has to be right. They stated that sufficient overlap in research interest has to exist between the partners (although not too much overlap so that no competition may develop), and that appropriate funding needs to be available or has to be mobilized. This would suggest the reason that time is an important factor for potential partners in inter-group collaboration to meet, to develop trust into the likely benefits of collaboration with one another, and to eventually carry through collaborative research. Nevertheless, even if we only regard the *continuous* cohort of clusters we still find the association of collaboration propensity with cluster size. For fields A and C this association is very pronounced, with only 13.5%, respectively 16.9% of small clusters collaborating, whereas 58.8%, respectively 62.7%, of the large clusters collaborate. For field B that shows a lower collaboration propensity overall, the corresponding numbers are 6.8% for small clusters, and 22.2% for large clusters. This lower involvement of small but continuous clusters in the collaboration network may indicate that their research focus is marginal to the research specialty under investigation, or that they contribute to the core knowledge of the specialty operating by a very specific (less-collaborative) working mode. Again, a better understanding of the social organisation and working mode of coauthor groups represented by small clusters should help to illuminate the reasons for their lower inter-group collaboration propensity.

Our results on the transfer and collaboration networks show that the extent of the giant components of our co-author networks is primarily defined by transfer type relationships between clusters. This implies that about 90% of authors with at least two publications in the field are part of a group network where groups are loosely in contact by some kind of transfer interaction described above. So whereas the transfer network can be characterized as a pervasive global cooperative network, the collaboration network is much more selective. In all three fields it includes less than a third of the clusters, and less than half of the authors from the giant components of the networks.

From the visualization in figure 8 of the dominant geographical affiliations of clusters in the collaboration network we can deduce that geographical proximity at continent level correlates with the local density of the collaborative network such that substructures (whether separate components in field B, or just denser subnetworks in fields A and C) tend to be geographically homogenous. Interestingly, (the much fewer) North-American clusters contrast somewhat with European and Asian clusters in that they do not form a distinct substructure, but seem more globally interspersed. Hence our analysis would seem to confirm the continentalization of science hypothesis of Leclerc & Gagner (1994) only for Europe, while North America and Asia expose a different pattern, at least for the research specialties under investigation here: North-American clusters have a rather low tendency to link to one another, and the Asian network exposes a distinct substructure of major national networks (based on author names we can identify a Chinese, Japanese, and a Korean sub-cluster in fields A and C, plus a small distinct Indian sub-cluster in field A).

**Mesoscopic Structure of Co-author Networks in Chemistry: Field Specific Differences**

Apropos of the differences that we observe between the three fields, there are some observations that point to unique field-specific conditions, such as the historical evolution of a field as reflected in the growth pattern of its co-author network, and the relative sizes of its cluster age cohorts. Also the features of the geographically resolved collaboration networks in figure 8 point to differences in the material conditions and research styles between the three fields. For example, the stronger integration of the network in field A may be the consequence of large, expensive, internationally shared instrumentation used in field A. The relative under representation of North American clusters especially in field C, and the lack of a dense North American sub-network of clusters, matches with the perception of one of our informants about funding priorities of the U.S. American science system that provides little funding for collaborative research in this research specialty.

We further observe that fields A and C have a number of features in common that contrast with properties of field B. These differences may well point to discipline specific working modes, since fields A and B are strongly affiliated with the same chemical sub-discipline, physical chemistry, whereas field B can be best characterized as synthetic chemistry.

- Co-author clusters in field B are smaller than in fields A, and C in the sense that field B has less large clusters, and more small clusters than fields A and C. This indicates that the size of basic collective units necessary to make relevant research contributions to the respective fields is smaller in synthetic chemistry than physical chemistry as represented by our fields A, B, and C.
- The predominance of single-hub clusters over multi-hub clusters in field B points to a preferred organizational model where one senior researcher based at an University leads a group of younger researchers and students that he or she trains. While this model is not unusual in fields A and C either, those fields show three times as many multi-hub clusters than field B, indicating a greater proportion research networks, either institutional or of distributed groups teaming up to form larger collective units.
- A greater collaboration propensity in fields A and C is further underlined by the analysis of the properties of the collaboration networks in the three fields. As figure 8 shows, the difference between the collaboration network in field B on the one hand, and the collaboration networks of fields A and C on the other hand, are striking. The small size of the collaboration network of field B is not a direct consequence of the smaller overall size of the co-author network of field B, but due to the lower percentage of clusters participating in the collaboration network, and the lower percentage of collaboration type between-cluster links. Also, the collaboration networks of field A and C are locally denser (higher cluster node degrees), and more cohesive, in the sense that they have a giant component, whereas the collaboration network of field B breaks up of into separate, disconnected components of similar sizes.
- Another indicator that points to a lower collaboration propensity of field B is the observed shift of the distribution of node role types for non-hubs and hubs in fields A and C towards node types that are more connected to other clusters. Field

B has a higher proportion of so called 'ultra-peripheral' non-hub nodes that do not have any outside coauthor links, as well as a higher proportion of 'provincial' hub nodes that have few if any outside links. This could be explained by field B having a majority of PI-led groups of students and possibly postdocs that have no or minimal collaborative links outside of their own group.

With only three research specialties in this case study the empirical basis is too small to generalize findings to the entire sub-disciplines of physical chemistry or synthetic chemistry. But the observations made suggest the sensitivity of our approach to detecting field-specific and possibly sub-discipline specific collaboration patterns, and they demonstrate its power to extract salient features and to generate directions for further research to understand the social configurations and processes underlying differences in collaboration patterns.

**Limitations of This Study**

Finally, we point to limitations of this study – some are technical that we plan to approach in future work, some are fundamental, inherent to the approach.

*Temporal Resolution*

We build the co-author networks from data covering a period of roughly 20 years. Our analysis introduces only a preliminary temporal analysis. We see some value in refining the temporal analysis to study cluster evolution with time, but there is an inherent limit to achieving this kind of resolution from publication data. Publications tend to be temporally sparse for indicating underlying research activity and social organisation (with sometimes many years' of delay until some result possibly obtained many years earlier gets published), such that if we chose a time window too small, we will underestimate the size of co-author groups. Hence to get a sense of a research group as a research performing collective and the kind of material and social resources that it builds upon, we need complementary information, e.g. from observation in a field study.

*Author Name Disambiguation*

Certain observations indicate that especially author name homonymy due to a small set of common surnames used widely in some East Asian countries such as China and Korea, distort the co-author networks and their structures. This issue will impact our analysis in several ways and the net effect requires careful study. From our ongoing work on such an analysis (to be published) we can report that the qualitative features discussed here, in particular the field differences in the collaboration networks including their geographical ordering remain valid also for a disambiguated version of the data sets used here.

*Field Delineation*

Another challenge remains the issue of field delineation and ensuring that a publication data set is representative for a research specialty. Depending on the field and the agreement among actors about its intellectual boundaries, developing appropriate lexical queries in interaction with our study participants is a time consuming, iterative process. In those areas we are studying there is no easy match to a defined set of specialised journals that one could focus on to short-circuit the process. In the current stage of our work we prefer this interactive process as it tells us also about the manifold perspectives of participants on their research context, and the complexity of the construct 'research specialty'. Still, there are ongoing efforts to refine and automate field delineation (Bassecoulard 2007, Mogoutov 2007) that might prove crucial for realizing larger comparative studies.

*Scope*

Our analytic method is calibrated to the research environment of certain fields in chemistry and physics. We believe that conceptually it can be transferred to other fields and disciplines as well. But it may have to be significantly altered to be effective, since in other research contexts different data and interaction forms will be relevant (e.g. in computer science the predominant role of conferences, and secondary role of journal publications). Finally, we are capturing here exclusively collaboration forms that manifest themselves in co-authorship - other forms of informal collaboration are not represented in the networks, although informants underline their importance to their work. There are additional traces of interaction and influence, such as inter-citation, or workshop participation, to complement the picture, but there will remain important forms of informal exchange that are not documented.

## Summary & Conclusions

Using a combination of qualitative and quantitative methods we arrive at fundamental insights about the collective organization of research specialties in chemistry. We base our study on observations from the mesoscopic analysis of linking patterns in clustered co-authorship networks. These point to different structural types of co-author groups as well as different types of collaborative relations between co-author groups. We clarify the meaning and validate the relevance of those structural differences by matching these structural features with real world scenarios described in interviews with field study participants. We proceed with an empirical investigation of such features in the co-authorship networks of three research specialties in chemistry. This exposes a number of generic features of collaboration such as its geographical ordering, as well as field-specific and possibly sub-discipline specific features such as dominant structures of the most basic collective units in a research specialty and inter-group collaboration propensity.

The empirical results presented in this paper generate further research questions, e.g.:
- What is the nature of the 'scatter of' small co-author groups – their social organization and scientific working mode; how does this scatter of small groups relate to the transient and less productive 'scatter authors' that Morris (2007, 2005) contrasts with the 'core authors' in a research specialty?
- How is international collaboration in a research specialty shaped; in what way and how strongly are national research communities internally and internationally interlinked? Whereas previous bibliometric studies into international collaboration based their analyses on the statistical co-occurrence of countries in paper affiliations (Glänzel & DeLange 1997, 2002, Zitt 2000), the analytic tools developed in this paper provide access specifically to the network of inter-group collaboration in a research specialty – and hence the self-organizing network character of science emphasized by Wagner & Leydesdorff (2005), and Leydesdorff & Wagner (2008), but at the level of research collectives in research specialties. Inspection of features of these networks will help to identify further locations of interest e.g. to explore motivations and conditions of collaboration.

We intend to further develop our methodology by a refinement of the network analysis to explore the temporal dimension of cluster growth, the structural organization of clusters as basic collective research performing units, and the role of author nodes in particular structural positions. An important pre-condition is a reasonably good disambiguation of author identities that we are currently working on. Further we plan to extend the empirical base into other research fields to explore variations in patterns and how they may relate to epistemic differences between fields.

**Acknowledgments**
We are indebted to our field study participants. Further, this research has been made possible through financial support by the National Science Foundation through grants IIS-738543 SGER: Advancing the State of eChemistry, DUE-0840744 NSDL Technical Network Services: A Cyberinfrastructure Platform for STEM Education, and NSF award 0404553. Support also came from Microsoft Corporation for the project ORE-based eChemistry. We are grateful to those that make our work so much more effective by making neat tools and algorithms available on the Web, such as Martin Rosvall (infomap clustering code), Vladimir Batagelj and Andrej Mrvar (pajek), Michael Weseman (plot), and Peter Mcaster (OmniGraffle extensions for pie charts).

# Appendix

## Appendix A: Data

Guided by our informants we developed lexical queries to retrieve from Web of Science the literature covering a specific speciality– sometimes a single term, sometimes a 2-3 line long query using Boolean operators. Our informants validated the data sets, checking whether crucial authors whom they expected to see were actually represented, and whether out-of-scope topics had been accidentally introduced. The data was then further processed to eliminate articles with only one author or unknown authors (indicated by 'Anon' in the WoS records), or with only a single reference, as the latter records typically do not refer to original research articles or reviews.

This study is part of a larger dissertation research project. To ensure the anonymity of the human participants of the field study part of this research in accordance with the IRB[vi] approved protocol for the dissertation research we cannot disclose the lexical queries that would provide access to the exact data sets discussed in this report. If access to the networks build from the data used in this study is needed for a particular research or validation purpose we are happy to work out a solution that retains the anonymity of our study participants.

Data Field A:
Time span: 1991-2008, number of records: 53,947 (retrieved on 22 Jun 2009)

Data Field B:
Time span: 1991-2008, number of records: 12,817 (retrieved on 14 Nov 2008)

Data Field C:
Time span: 1987-2008, number of records: 30,636 (retrieved on 17 Dec 2008).

We derive the geographical affiliation of a cluster at continent level from the country affiliation listed in the Web of Science record for each publication. We represent each cluster by all the publications any of its authors has been a co-author of. Then we determine the country affiliation that is most often listed for papers published by cluster authors. In cases where the second placed country is listed at least 50% as many times as the most often listed country, and if these two countries belong to different continents, we assign a mixed, two continent geographical affiliation

**Appendix B: Methods**

In Guimera et al. (2007) each node of a clustered network was assigned to one out of seven *node roles*. The seven node roles are defined by the values of two parameters that quantify how a node is connected to the other nodes in its cluster and how the links going outside the cluster are distributed.

The first parameter, z, is simply the inside-its-cluster-degree of a node, normalized by subtracting the average inside-the-cluster degree and then dividing the result by the standard deviation. Nodes that have z < 2.5 were called *non-hubs*, while nodes with z ≥ 2.5 were called *hubs*. In other words, hubs are well-connected inside their own cluster with a normalized threshold beyond 2.5 (z > 2.5).

The second parameter is the participation coefficient P that quantifies how a node distributes its outside links among the clusters. For a node *i* in a network of *N* clusters it is defined as follows:

$$P_i = 1 - \sum_s (k_s^i / k^i)$$

where s=(1….*N*),

$k_s^i$: number of links of node *i* to nodes in cluster s,

$k^i$: the total degree of node *i*.

From this definition it follows that P is normalized for each node so that the value is within 0 and 1. A node that connects to only one cluster would have a small P whereas a node that connects to a lot of clusters would have a high P.

Non-hubs are classified into 4 types based on their P values.
- P ≤ 0.05 : *Ultra-peripheral* (R1) with its connections restricted to its own cluster
- 0.05 < P ≤ 0.62 : *Peripheral* (R2) with limited outside connectivity
- 0.62 < P ≤ 0.8 : *Satellite connector* (R3) with good connectivity with a number of outside clusters
- 0.8 < P : *Kinless* (R4) with outside connectivity distributed evenly among all clusters

Similarly hubs are categorized into 3 types based on their P values.
- P ≤ 0.30 : *Provincial hub* (R5) that is well-connected only inside its own cluster
- 0.30 < P ≤ 0.75 : *Connector hub* (R6) that connects to many clusters
- 0.75 < P : *Global hub* (R7) that connect to most clusters

We have used the same set of threshold values to assign roles to nodes.

[i] We call these 'seed' groups as we regard them as entry points into scientific communities, and plan to extend our field studies following links (of cooperation or competition) of these seed labs.

[ii] It is worth noting that these specialties do not fall squarely into single sub-disciplines but typically unite the efforts of researchers who identity with different subdisciplines. For field B which we label here as 'synthetic chemistry' these are mainly organic chemists, inorganic and organo-metallic chemists as well as polymer chemists. For field A which we label here as belonging to the subdiscipline 'physical chemistry' these are physical chemists as well as experimental and theoretical physicist, often with a background in atomic and molecular physics, but also in nuclear physics. For field C these are indeed mostly physical chemists.

[iii] When comparing this number to other co-author networks in the literature, remember that this number is calculated for a reduced co-author network (after excluding one-time authors). Hence this number will overestimate the relative size of the giant component for the unreduced network.

[iv] Available from Martin Rosvall's home page at http://www.tp.umu.se/~rosvall/code.html

[v] Our field studies confirm that in areas of synthetic chemistry the task of having to find someone with the instrumental equipment to conduct certain measurements on your sample is very common.

[vi] Institutional Review Board, http://en.wikipedia.org/wiki/Institutional_review_board

# Supplementary Material for the Article:
# "A New Approach to Analyzing Patterns of Collaboration in Co-authorship Networks - Mesoscopic Analysis and Interpretation"

Theresa Velden, Asif-ul Haque, and Carl Lagoze

*tav6, ah344, cjl2 @cornell.edu*
Computer and Information Science, Cornell University

Note: in addition to this text file with supplementary information, we are also providing the following data files:

1. Co-author network files .net in pajek format for fields A, B, C
2. .csv file with cluster properties for fields A,B, C

**Table of Contents:**

# 1. Distribution of Cluster Sizes

The cluster size distributions of all three fields are strongly right skewed, and non-Gaussian (figure S1). Especially field B shows a peculiar, flattened distribution which may be due to the rather small sample size (578 clusters). If one inspects the far right tail of the distributions (cluster sizes > 40) one can see that only a very small fraction of clusters in B is very large. This is the most striking difference between fields A, and C, versus field B. If one looks at the left part of the diagram with cluster sizes ≤ 40, then the differences between the fields are less pronounced – this is reflected in the percentiles given in table 1. The distribution of the smallest 25% of clusters is very similar for all three fields. Only the median gives a first indication that clusters in fields A and C are further shifted towards larger sizes than in field B.

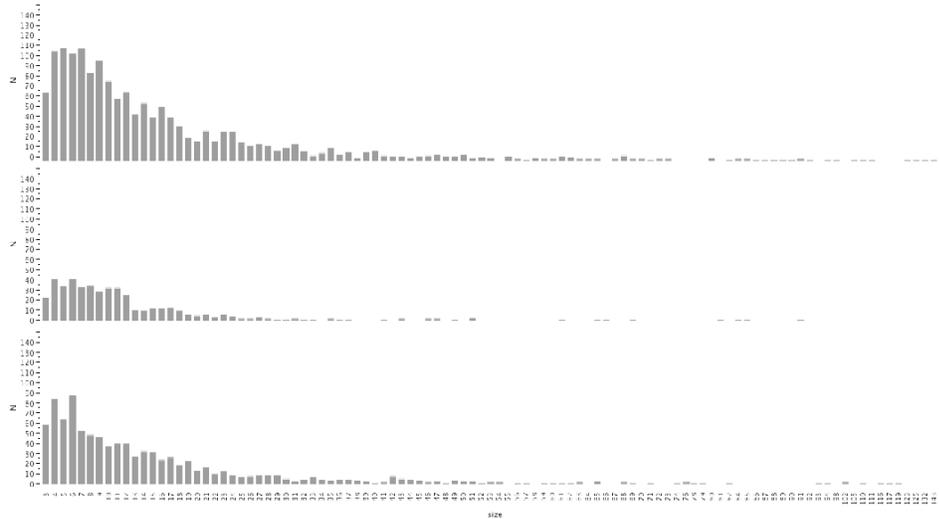

**Figure S1: Size distributions of clusters in giant component of fields A, B, and C.**

|  | Minimum | 10% | 25% | Median | 75% | 90% | Maximum |
|---|---|---|---|---|---|---|---|
| **Field A** | 3 | 4 | 7 | 11 | 20 | 35 | 143 |
| **Field B** | 3 | 4 | 6 | 9 | 14 | 22 | 91 |
| **Field C** | 3 | 4 | 6 | 10 | 18 | 32.8 | 159 |

**Table S1: Percentiles for cluster sizes in giant component of fields A, B, and C.**

## 2. Cluster Hubness & Cluster Size

The three scatter plots in figure S2 show nicely the 'stacking' of the hubness property with clusters size for all three fields.

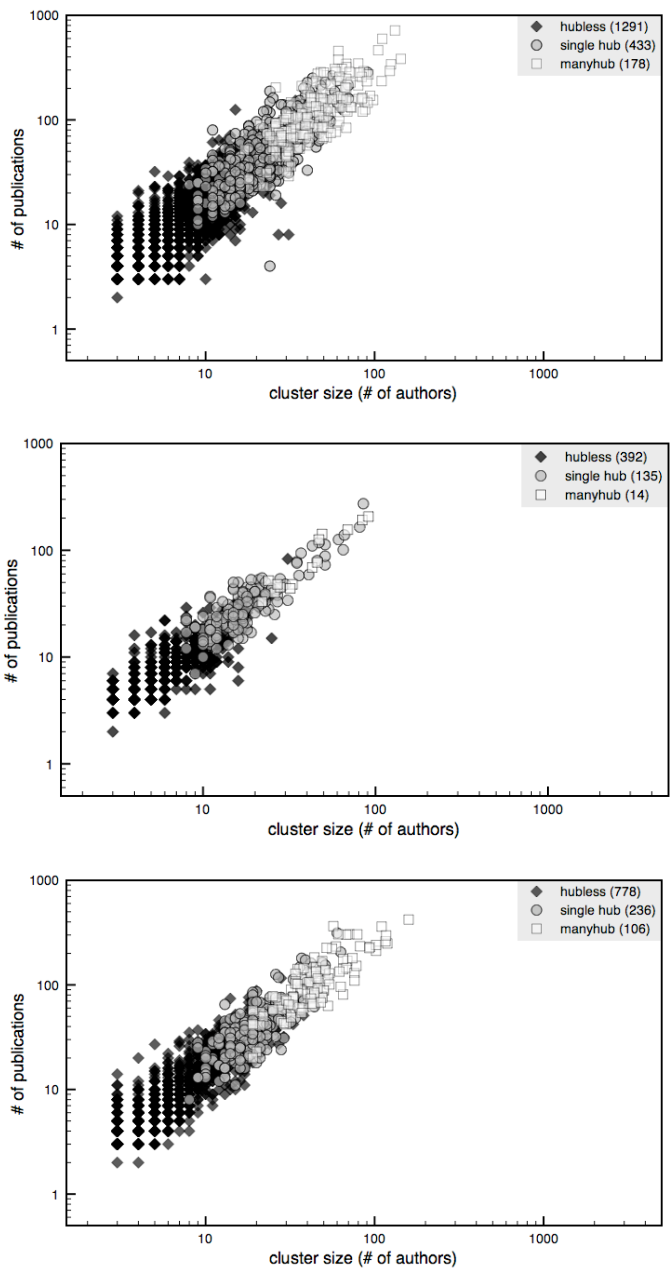

**Figure S2: scatter plots for fields A (top), B (middle), and C (bottom) showing cluster size vs. number of publications of cluster authors in the data set. The hubness of each cluster is indicated by the following symbols: white square: *multi-hub*, grey circle = *single hub*, black diamond = *no hub*.**

We find across all three fields a statistically significant association between the categorical variables cluster size and cluster hubness (N=3564, DF=4, Pearson ChiSquare = 2558.8, Prob > ChiSq: = 0.0000). The mosaic plot in figure S3 shows how each value of the hubness variable is distributed across the cluster size variable.

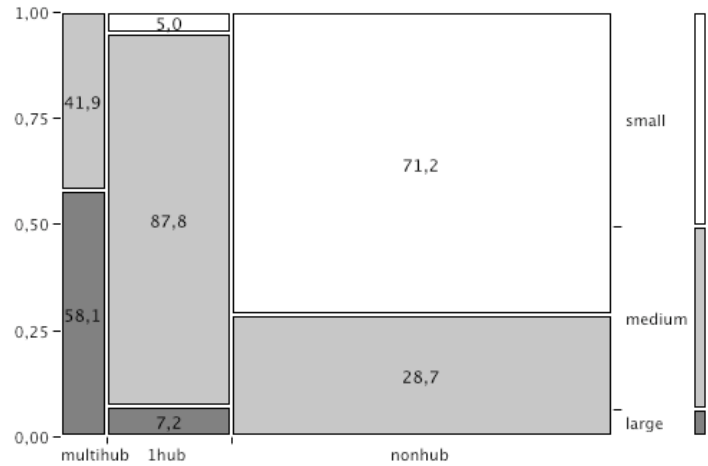

**Figure S3: Mosaic plot showing for the aggregate set of clusters from all three fields percentages of small, medium, and large clusters for clusters of different hubness.**

## 3. Cluster Hubness & Field

Statistical testing confirms a significant association of the distribution of the cluster hubness variable over the three fields (N=3564, DF=4, Pearson ChiSquare = 29.151, Prob > ChiSq: < 0.0001).

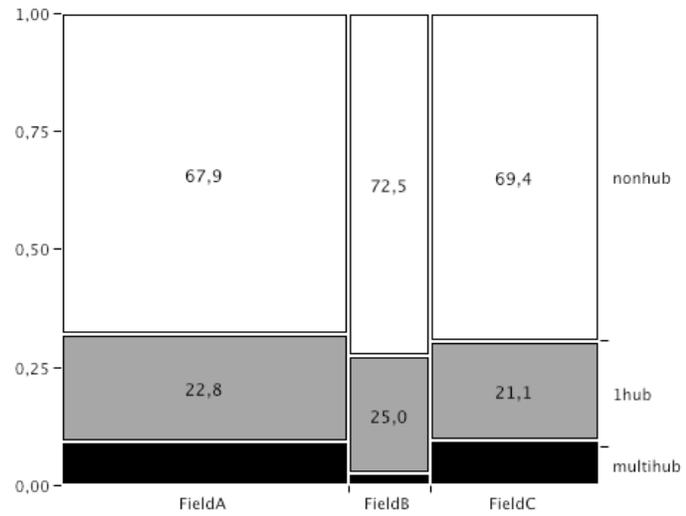

**Figure S4: Mosaic plot showing for the distribution of cluster hubness over the different fields. Numbers are given in percentage of clusters in the respective field.**

## 4. Age Distribution of Small, Hubless Clusters

The diagrams in figure S5 show the age distributions of the small, hubless clusters for each field. We see that the large majority of small, hubless clusters are either in the continuous or the recent age cohort – this means that some of their authors have been publishing in the field during the last two or even all three time slices.

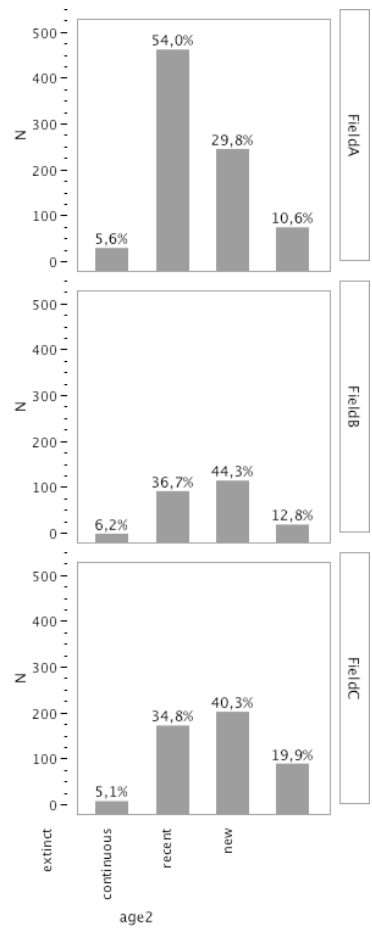

**Figure S5: size distribution of the subset of small, hubless clusters for each field.**

## 5. Properties of Collaborating Clusters: Age and Size

As can be seen from the mosaic plots shown in figure S6, for each of the three fields holds that collaborating clusters tend to be older and larger than non-collaborating clusters. Figure S7 shows that in fields A and C a larger fraction of large clusters is collaborating than in field B.

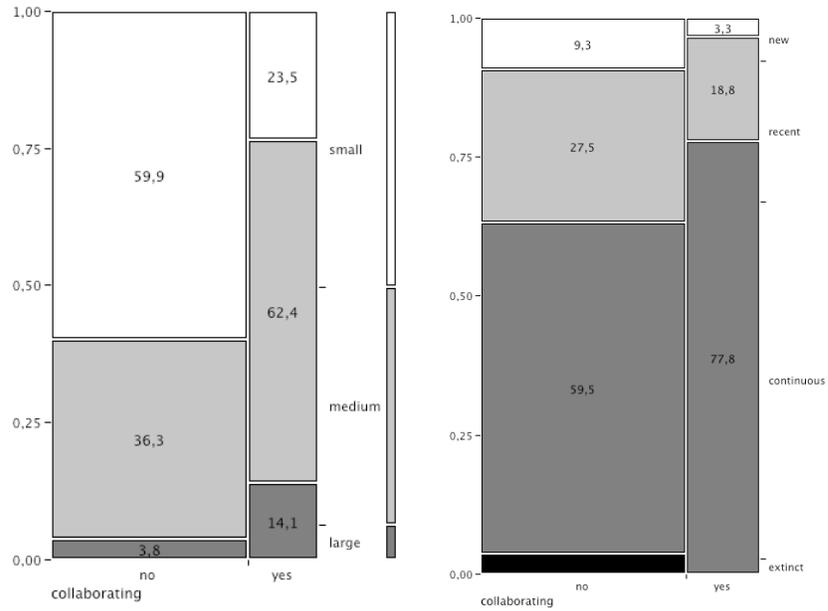

**Figure S6: Mosaic plots showing for the distribution of cluster size (left), and cluster age (right) for collaborating vs. non-collaborating clusters. Numbers are given in percent of collaborating, respectively non-collaborating clusters.**

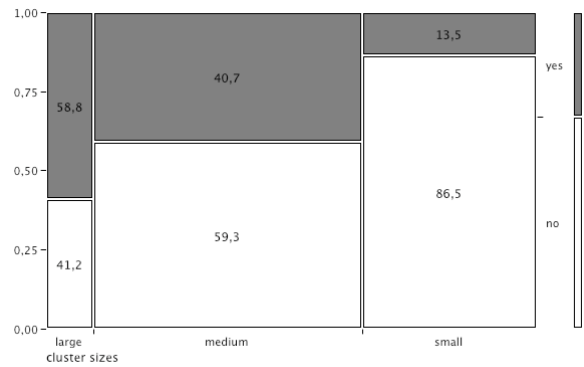

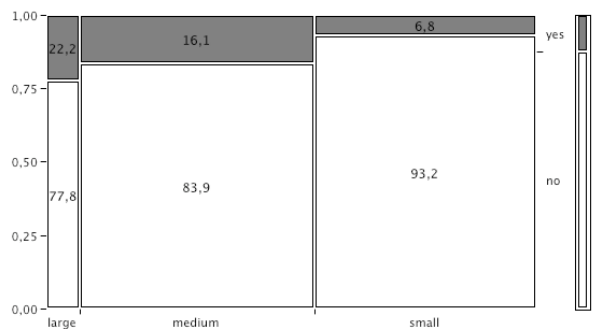

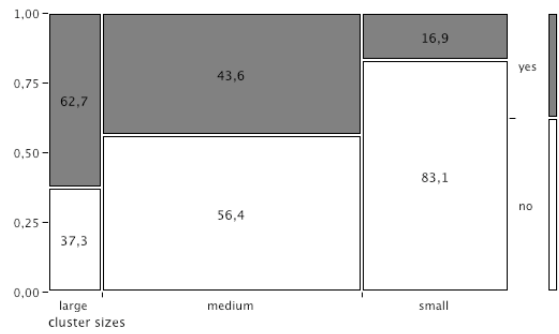

**Figure S7: Mosaic plots showing for the subset of continuous clusters the distribution of cluster sizes for collaborating vs. non-collaborating clusters. Top: field A, middle: field B, bottom: field C. Number are given in percent of collaborating, respectively non-collaborating clusters.**